# Hierarchical Temporal and Spatial Clustering of Uncertain and Time-varying Load Models

Xinran Zhang, *Member, IEEE,* David J. Hill, *Life Fellow, IEEE*

*Abstract*—Load modeling is difficult due to its uncertain and time-varying properties. Through the recently proposed ambient signals load modeling approach, these properties can be more frequently tracked. However, the large dataset of load modeling results becomes a new problem. In this paper, a hierarchical temporal and spatial clustering method of load models is proposed, after which the large size load model dataset can be represented by several representative load models (RLMs). In the temporal clustering stage, the RLMs of one load bus are picked up through clustering to represent all the load models of the load bus at different time. In the spatial clustering stage, the RLMs of all the load buses form a new set and the RLMs of the system are picked up through spatial clustering. In this way, the large sets of load models are represented by a small number of RLMs, through which the storage space of the load models is significantly reduced. The validation results in IEEE 39 bus system have shown that the simulation accuracy can still be maintained after replacing the load models with the RLMs. In this way, the effectiveness of the proposed hierarchical clustering framework is validated.

*Index Terms*-- Load modeling, clustering, power system uncertainty.

## I. INTRODUCTION

DUE to the uncertain and time-varying properties of power loads, load uncertainty is an important cause of power system uncertainty. Load uncertainty can be further divided into two categories, i.e. load amount uncertainty [1] and load model uncertainty [2]. Load model uncertainty was considered by changing the proportions of different load components in [2]-[4], after which the impacts of load model uncertainty on power system transient stability and damping control are studied.

However, the data of load model uncertainty is randomly generated without any basis in previous research. How to describe load model uncertainty based on practical measurement data has not been considered. An important reason is that previously it is not easy to follow the time-varying and uncertain changes of load models from practical measurements. In previous research of load modeling, measurement-based load modeling is mainly based on post large disturbance response (PLDR) data [5], [6]. However, the limitation of this approach is its dependence on the occurrence of large disturbance events. For the periods without large disturbance events, load models cannot be built. Therefore, the time-varying and uncertain changes of load models cannot be tracked by PLDR based approach.

The recently proposed ambient signals-based load modeling approach provides a promising method to track the changes of load models [7]. Ambient signals in the measurement data of power and voltage are caused by the continuous random changes in loads and renewables in power systems. Then, by analyzing the relationship between power ambient signals and voltage ambient signals, load model parameters can be identified accordingly. This approach makes it possible to periodically track the time-varying and uncertain changes of load models. In this way, load model uncertainty can be considered in power system dynamic analysis. Apart from ambient signals based approach, the robust time-varying load modeling approach can also be applied to track load model uncertainties [8], [9].

Nevertheless, a new problem is raised that the dataset of load models becomes large in periodical ambient signals-based load modeling. For example, if ambient signals-based load modeling is conducted every 15 minutes, altogether 96 load models will be built for one load bus in one day. As a result, the storage cost of load modeling results will increase significantly. In addition, when performing power system simulation based on load modeling results, it is also difficult to select which load models in the dataset should be applied. Therefore, it is necessary to further reduce the number of recorded load models.

After analyzing the necessity of reducing the number of recorded load models, the next problem is the feasibility. Although the total number of load models obtained from ambient signal-based identification is very large, many elements from the load models' set may be similar. In addition, the changes in load models are mainly caused by the changing behavior of the power consumers. Then, similar load models may appear repeatedly at different time. Therefore, it is reasonable to assume that the load models can be grouped into several clusters, with all the load models in one cluster being similar. This assumption can be supported by the conclusion in [10], which shows that load models' characteristic can be captured and modeled in a smaller subspace of the data space.

Clustering methodology, which belongs to the scope of unsupervised machine learning, provides a promising way to reduce the number of load models. Through clustering, similar elements in a set can be grouped in to one cluster, and a cluster center can be picked up to represent all the elements in the same cluster. In this way, the cluster centers, which are also known as representative elements in the set, are picked up to represent all the elements in the set. It has been applied in the research of power system in many aspects, including wind farm model aggregation [11], and energy consumption behavior [12]. In this paper, clustering is applied to load models so that representative load models (RLMs) can be picked up to represent all the load models. In this way, the number of load models is reduced to the number of RLMs. In previous research, load model clustering is firstly proposed in [13]. Load models are clustered through K-medois algorithm, in which the distance of load models is defined by post-fault response (PFR).

Based on the PFR based distance of load models in our previous work [13], a hierarchical clustering framework of load models is designed in this paper. Two stages are included in the hierarchical framework, i.e. temporal clustering, and spatial clustering. The identification results of load models are processed by the proposed hierarchical clustering framework, after which the number of load models is significantly reduced without deteriorate the accuracy of load model dynamic performance. Compared with our previous work, the following improvements are made. Firstly, a more complicated load model structure is adopted. Secondly, a better density based clustering algorithm is applied, which can provide more robust clustering results without iteration. Thirdly, the hierarchical framework is newly designed, which can further reduce the number of load models.

The rest of this paper is organized as follows. In section II the load model structure and the identification of load model parameters are introduced. Section III proposes the hierarchical load model clustering framework. Section IV introduces the distance between load models and the clustering algorithm. Section V presents the case study results in IEEE 39 bus system. Section VI concludes this paper.

## II. LOAD MODEL STRUCTURE AND IDENTIFICATION

### A. Composite Load Model Structure

The widely used composite load model is used as the load model structure in this paper. The composite load model consists of two parts, i.e. a static part and a dynamic part. In this section, the model structure and the parameters of these two parts are introduced. A group of parameters of composite load model ($Pa$) includes the following parameters, the proportion of dynamic load $p$, the active static load model parameters $Pas$, the reactive static load model parameters $Prs$, and the dynamic load model parameters $Pd$.

*1) Dynamic Load Model: Induction Motor*

In a dynamic load model, power consumption is related not only to the current voltage synchrophasor but also to the past model states. Therefore, the relationship between power consumption and voltage is described by a state space formulae model, which includes the state formulae and the output formulae. The third-order model of the induction motor is used in this paper to represent the dynamic load, the state formulae and output formulae of which are given as follows:

$$\begin{cases} \dfrac{dE_d}{dt} = -\dfrac{X}{T_{d0}X'}E_d + s\omega_0 E_q + \dfrac{1}{T_{d0}}(\dfrac{X}{X'}-1)U_d \\ \dfrac{dE_q}{dt} = -\dfrac{X}{T_{d0}X'}E_q - s\omega_0 E_d + \dfrac{1}{T_{d0}}(\dfrac{X}{X'}-1)U_q \\ \dfrac{ds}{dt} = \dfrac{1}{H_2}(T_m - \dfrac{E_d U_q - E_q U_d}{X'}) \end{cases} \quad (1)$$

$$\begin{cases} P_d = \dfrac{E_d U_q - E_q U_d}{X'} \\ Q_d = \dfrac{U_d^2 + U_q^2 - U_d E_d - U_q E_q}{X'} \end{cases} \quad (2)$$

where $U_d$ is the d-axis component of the voltage synchrophasor, $U_q$ is the q-axis component of the voltage synchrophasor, $X$ is the rotor open circuit reactance, $X'$ is the rotor transient reactance, $T_{d0}$ is the rotor open-circuit time constant, $H_2$ is the inertia time constant, $\omega_0$ is the synchronous rotation angular speed, and $T_m$ is the mechanical torque. Therefore, there are altogether 5 parameters to be identified, i.e. $Pd=[X, X', T_{d0}, H_2, T_m]$.

*2) Static Load Model: ZIP*

In a static load model, power consumption is only related to current voltage magnitude ($U$), which can be expressed by algebraic formulae. In this paper, the ZIP model is selected as the static load model. The relationship between power consumption and voltage magnitude of the ZIP model is given as follows,

$$\begin{cases} P_s = P_z U^2 + P_i U + P_p \\ Q_s = Q_z U^2 + Q_i U + Q_p \end{cases} \quad (3)$$

where $P_s$ is the active power consumption, $Q_s$ is the reactive power consumption, $U$ is the current voltage magnitude, $Pas=[P_z\ P_i\ P_p]$ and $Prs=[Q_z\ Q_i\ Q_p]$ are static load model parameters.

### B. Ambient Signals based Load Modeling

Due to the constant random changes in power loads and renewables, ambient signals are always existing in power system voltage and power signals. Then, by analyzing the relationship between voltage ambient signals and power ambient signals of load buses, the load models of these buses can be identified accordingly.

In practical situations, ambient signals-based load modeling can be conducted periodically, e.g. every 15 minutes, to better track the time-varying changes of load models. Within one identification period, the load model of one bus are assumed not to obviously change, therefore it can be represented by the identification results at the end of each period. A small piece of measurement data is selected for load model parameter identification at the end of each period, the length of which is much shorter than the identification period (e.g. 10s). The aim of selecting the small pieces of data is to improve computation efficiency. The whole timeline is illustrated in Fig. 1, in which $n$ times of load model identification are conducted periodically.

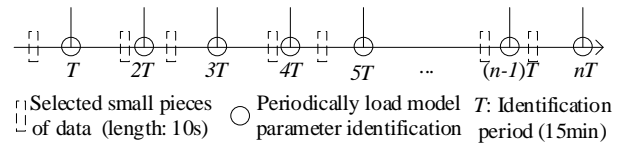

Fig. 1 Timeline of ambient signals-based load modeling

## III. HIERARCHICAL CLUSTERING FRAMEWORK OF LOAD MODELS

### A. Hierarchical Clustering Framework

Through ambient signals-based load modeling, the load models of one load bus at different time can be obtained periodically, after which a set of load models can be formed. Define $\mathbf{B}=\{1, 2, …, m\}$ as the set of load buses, and $\mathbf{T}=\{1, 2, …, n\}$ as the set of time when the load models are identified. In this way, the load model of Bus $i$ identified at time $j$ is recorded as $Pa_{ij}$, $i\in \mathbf{B}$ and $j\in \mathbf{T}$.

In this section, the hierarchical clustering framework of picking up RLMs from load models' set is proposed. Fig. 2 shows the timeline of load model identification and clustering in multiple load buses. The hierarchical clustering framework mainly includes two stages, i.e. temporal clustering, and spatial clustering. Temporal clustering is to pick up RLMs of one load bus from the load models' set of the bus, which is conducted locally at each load bus. Spatial clustering is to pick up RLMs of the system from the RLMs' set of all load buses in the system, which is conducted at the control center.

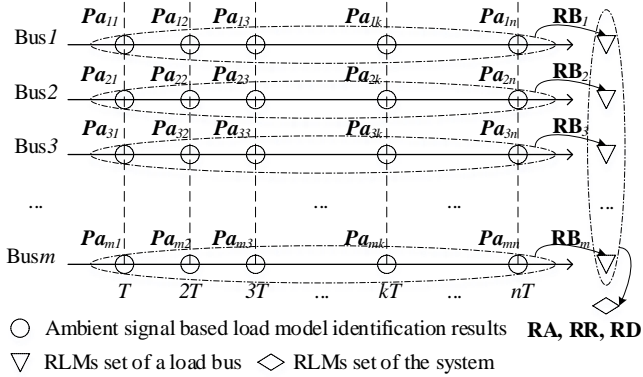

Fig. 2 Hierarchical clustering framework of load models

*B. Temporal Clustering*

Through periodical ambient signals-based load model parameter identification, a set of different load models is obtained in each load bus, as shown in Fig. 2. The changes in the load models of one load bus are mainly caused by the behavior of the power consumers around the load bus, including the on/off behavior and the change of power consumption amount. The behavior of power consumers should follow some regular periodical patterns at different time. Therefore, it is reasonable to assume that load models of one load bus at different time can be grouped into several clusters, with similar load models in the same cluster.

Temporal load model clustering is conducted locally at each load bus, based on the ambient signals-based load model identification results. The task of temporal load model clustering is to group similar load models into clusters and pick up one RLM for each cluster. The set of load models of bus $i$, $i \in \mathbf{B}$ includes $Pa_{ij}$, $j \in \mathbf{T}$. Through temporal load model clustering, the RLMs of Bus $i$ are picked up, the set of which are noted as $\mathbf{RB}_i$, $i \in \mathbf{B}$. The items in $\mathbf{RB}_i$ are noted as $rb_{ik}$, $k=1, 2, \ldots, r_i$, where $r_i$ is the number of RLMs of Bus $i$. $rb$ has the same form with $Pa$. Then, in an RLM $rb_{ik}$, the proportion of dynamic load is noted as $rp_{ik}$, the set of active static load model parameters is noted as $rba_{ik}$, the set of reactive static load model parameters is noted as $rbr_{ik}$, and the set of dynamic load model parameters is noted as $rbd_{ik}$.

Afterwards, only the sets of RLMs $\mathbf{RB}_i$, $i \in \mathbf{B}$ are recorded and sent to the control center, the number of which is much smaller than the total number of load models at the load buses.

*C. Spatial Clustering*

After the periodical identification and temporal clustering of load models at load buses, the RLMs of different load buses are sent to the control center for further processing. As there are many load buses in a power system, the total number of the RLMs of all the load buses is still very large. Therefore, further clustering aiming at picking up the RLMs of the whole power system is necessary, which is called spatial clustering. Then, the number of load models to be stored can be further reduced.

The spatial load model clustering is conducted at the control center of a power grid. The task of spatial clustering is to pick up representative active static load models (RASLMs), representative reactive static load models (RRSLMs), and representative dynamic load models (RDLMs) of the power system separately. After receiving all the $rba_{ik}$, $rbr_{ik}$ and $rbd_{ik}$, they are rearranged as $\mathbf{RBA}=\{rba_l\}$, $\mathbf{RBR}=\{rbr_l\}$ and $\mathbf{RBD}=\{rbd_l\}$, $l \in \mathbf{L}$, $\mathbf{L}=[1, 2, \ldots, \Sigma r_i\,(i \in \mathbf{B})]$.

Firstly, the RASLMs are picked up from $\mathbf{RBA}$, the set of which is noted as $\mathbf{RA}$. The items in $\mathbf{RA}$ are noted as $ra_e$, $e=1, 2, \ldots, na$, where $na$ is the number of RASLMs. Secondly, the RRSLMs are picked up from $\mathbf{RBR}$, the set of which is noted as $\mathbf{RR}$. The items in $\mathbf{RR}$ are noted as $rr_f$, $f=1, 2, \ldots, nr$, where $nr$ is the number of RRSLMs. Thirdly, the RDLMs are picked up from $\mathbf{RBD}$, the set of which is noted as $\mathbf{RD}$. The items in $\mathbf{RD}$ are noted as $rd_g$, $g=1, 2, \ldots, nd$, where $nd$ is the number of RDLMs. Finally, three mapping indexes are given to each $rb_{ik}$ to show $rba_{ik}$, $rbr_{ik}$ and $rbd_{ik}$ are represented by which $ra_e$, $rr_f$ and $rd_g$, respectively. These three indexes for $rb_{ik}$ are noted as $ia_{ik}$, $ir_{ik}$ and $id_{ik}$, respectively. Therefore, $ia_{ik} \in \{1, 2, \ldots, na\}$, $ir_{ik} \in \{1, 2, \ldots, nr\}$, and $id_{ik} \in \{1, 2, \ldots, nd\}$. After spatial clustering, $rb_{ik}$ is simplified as $rb^{\sim}_{ik}=[rp_{ik}\ ia_{ik}\ ir_{ik}\ id_{ik}]$, which includes one dynamic load proportion parameter and three mapping indexes. In the rest of the paper, the same numbers are selected for $na$, $nr$ and $nd$, which is noted as $nc$. The dataflow of temporal and spatial clustering is illustrated in Fig. 3.

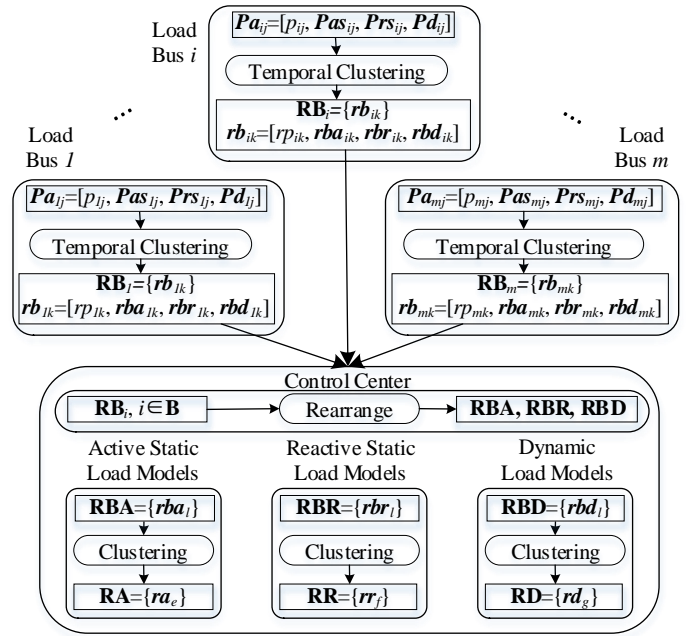

Fig. 3 Dataflow of hierarchical load model clustering

IV. CLUSTERING PROCEDURES AND ALGORITHM

*A. Distance between Load Models*

In clustering process, similar elements are grouped into one cluster, and then a representative element (which is also known as the cluster center) is selected to represent all the elements in the cluster. The similarity between elements is measured by the distance between elements. Therefore, in load model clustering, the distance between load models should be firstly defined.

As the application scenario of clustering results is power system dynamic simulation, the distance between load models should be related to the dynamic performance of different load models. The load models with similar dynamic performance after the same disturbance event can be grouped into one cluster in this way. Then, when applied in power system dynamic simulation, the RLMs can be used to represent other load models in the same cluster to get similar simulation results.

Therefore, the distance between load models in this paper is defined as the sum of the Euclidean distance between multiple post-fault response (PFR) curves of the load models. To begin with, the PFR curves of one load model are generated in a small-scale test system under multiple faults with various fault depth. Two PFR curves are generated for one load model, i.e. an active power curve and a reactive power curve. For the group of load model parameters $Pa$, the time series PFR curves under different faults are noted as PF($Pa$)={$y_1(Pa, t, p)$, $y_1(Pa, t, q)$, $y_2(Pa, t, p)$, $y_2(Pa, t, q)$, ..., $y_h(Pa, t, p)$, $y_h(Pa, t, q)$,}, where $h$ is the number of faults being tested. Then, the distance $d_{ij}$ between two load models $Pa_i$ and $Pa_j$ is defined as follows:

$$d_{ij} = \sum_{k=1}^{h}\sum_{t}(y_k(Pa_i,t,p) - y_k(Pa_j,t,p))^2 + (y_k(Pa_i,t,q) - (y_k(Pa_j,t,q))^2 \quad (4)$$

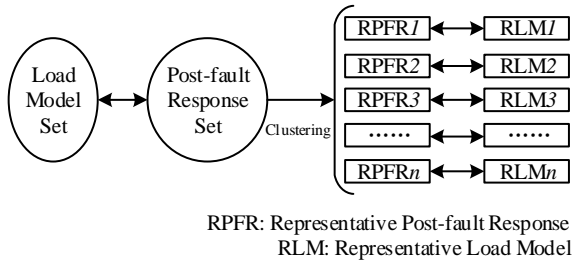

RPFR: Representative Post-fault Response
RLM: Representative Load Model

Fig. 4 Clustering of load models through post-fault response

In this way, a one to one mapping between load models' set and PFR curves' set is built. Each group of load model parameters $Pa$ is mapped to a group of PFR curves PF($Pa$). The distance between two groups of PFR curves is the same to the distance between two load models. In the clustering process, the representative post-fault response (RPFR) curves are selected from the PFR curves' set, and the corresponding load models for the representative post-fault response curves are then selected as the representative load models (RLM). This process is shown in Fig. 4.

### B. Post-fault Response Generation

*1) Basic Method*

As mentioned in the previous section, the PFR curves of the load models are generated in a small-scale test system. In this process, one of the load buses in the test system is selected as the test bus. With the load consumption of the test bus being unchanged, different load models are connected to the test bus, after which the PFR curves under multiple faults with various depth can be generated.

Different load model components are clustered at different clustering stages. In temporal clustering stage, the whole composite load model with a static load part and a dynamic load part is considered together in clustering. In spatial clustering stage, dynamic load model, active static load model and reactive static load model are clustered separately. Therefore, in the following sections, the methods to generate PFR at different clustering stages are discussed respectively.

*2) PFR Generation in Temporal Clustering*

The task of temporal clustering is to pick up $rb_{ik}$ from $Pa_{ij}$, $i \in \mathbf{B}$, $j \in \mathbf{T}$ and $k=1, 2, ..., r_i$. Since $Pa_{ij}$ includes both static and dynamic load model parameters, the load model which is connected at the test bus of the test system also includes these two parts. The proportion of dynamic load is $p_{ij}$. Then, through PFR clustering and the mapping from RPFRs to RLMs, $rb_{ik}$ can be picked up.

*3) PFR Generation in Spatial Clustering*

In spatial clustering, **RA**, **RR** and **RD** are picked up separately. Then, the PFR generation of active static load models, reactive static load models and dynamic load models is also conducted separately. In different process, the load models which are connected at the test bus are also different.

When generating the PFR curves of static load models to pick up **RA** and **RR**, only static load model is connected at the test bus. When generating the PFR curves of different active static load models, the reactive static load model is set to be constant Q. Similarly, when generating the PFR curves of different reactive static load models, the active static load model is set to be constant P. In this way, **RA** and **RR** can be picked up respectively.

When generating the PFR curves of dynamic load models to pick up **RD**, only dynamic load model is connected at the test bus. Then, the PFR curves of the dynamic load models is used in clustering to pick up RPFRs and the RLMs.

### C. Clustering Algorithm

After defining the distance between load models, the next step is to choose the clustering algorithm. Since the RPFRs picked up from the post-fault response set are to be mapped back to the RLMs according to the process in Fig. 4, only the clustering algorithm with the original elements as the cluster centers can be applied in solving the load model clustering problem. In this paper, a fast search and density peaks-based clustering (FDC) algorithm is applied [14].

In clustering problems, an element with more "neighbors" is regarded to have a larger density, in which the "neighbors" can be defined according to a cutoff distance $d_c$. The basic idea of this clustering algorithm is that the cluster centers are characterized by a higher density than their neighbors and by a relatively large distance from the points with higher densities. Therefore, for each element $i$, two properties are calculated from the distances among the objects, i.e. the density $\rho$ and the distance to the nearest point with higher density $\delta$.

The first step of the algorithm is to calculate the density of each element. The density of object $i$, which is defined as $\rho_i$, is calculated as follows,

$$\rho_i = \sum_{j}\gamma(d_{ij} - d_c) \quad (5)$$

where $d_c$ is the cutoff distance and $\gamma(x)$ is a bool function defined as follows: $\gamma(x)=1$ if $x>0$ and $\gamma(x)=0$ otherwise. One can choose $d_c$ so that the average number of the neighbors is 1-2% of the total number of points in the data set.

The second step is to calculate the distance to the nearest point with higher density, which is defined as $\delta_i$ for element $i$. If element $i$ is not the one with the largest density, $\delta_i$ is calculated as follows,

$$\delta_{i \neq \arg\max(\rho)} = \min_{j:\rho_j > \rho_i}(d_{ij}) \quad (6)$$

For element $i$ with the largest density, which means there is no element with higher density, $\delta_i$ is defined as the largest distance between object $i$ and the elements in the set, as follows,

$$\delta_{i = \arg\max(\rho)} = \max_j(d_{ij}) \quad (7)$$

The third step is to select the cluster centers accordingly. A decision map can be formed with $\rho$ being the $x$-axis and $\delta$ being the $y$-axis. The number of clusters are decided manually, after which the points with larger $\rho\delta$ are chosen as cluster centers.

Finally, the last step is to assign which cluster the remaining elements belong to. Each remaining element is assigned to the same cluster as its nearest neighbor of higher density. In this way, the clustering process is completed and each element is assigned to one cluster.

The points with large $\delta$ but small $\rho$ are the single points which are far away from the cluster centers and are regarded as outliers. In this paper, outliers are judged as follows: for one element, if its $\rho<0.001$ and its $\delta$ is larger than the average $\delta$ of all the cluster centers, it will be regarded as an outlier. The load models represented by the outliers will also be regarded as an RLM. In addition, the elements whose nearest point with higher density is an outlier will be assigned to the same cluster as the outlier. In this way, the number of clusters after clustering may be more than the initial choice due to the impact of outliers.

## V. CASE STUDY

### A. Introduction of Power Systems in Case Study

*1) Simulation System*

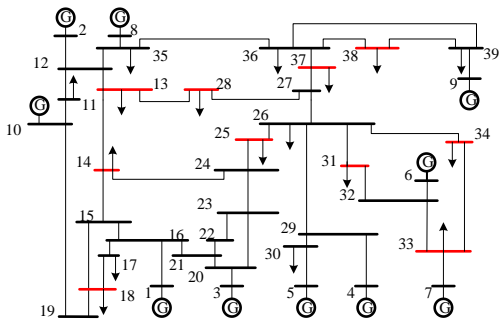

Fig. 5 Structure of IEEE 39 bus system

The IEEE 39 bus system is selected as the simulation system in this paper, in which there are 9 generator buses and 17 load buses. Among all the load buses, the models for 10 of them are selected to be composite load models, which are the test load buses considered in hierarchical load model clustering. These test load buses are marked in red in Fig. 5. The load models of these load buses are assumed to be already obtained from ambient signals based identification, and 500 different load models at each bus are recorded for further clustering. The load models of the other load buses are constant Z load models. The simulation cases in this section are conducted in MATLAB Power System Analysis Toolbox. The time step in simulation is 0.01s and the base value of system capacity is 100MVA.

*2) Test System to Generate Post Fault Response Curves*

The test system to generate PFR curves is selected to be the WSCC 9 bus system, the structure of which is given in Fig. 6. There are 3 generator buses and 3 load buses in this system. The test bus to connect different load models to generate PFR curves is chosen as Bus9. The load models of the other two load buses are constant PQ models. Three different faults are considered in PFR curves generation, i.e. the three-phase to ground fault at Bus7, Bus4 and Bus8. The voltage curves of these three faults with the same load models connected at Bus9 is given in Fig. 7. It can be observed that the depth of these three faults is different.

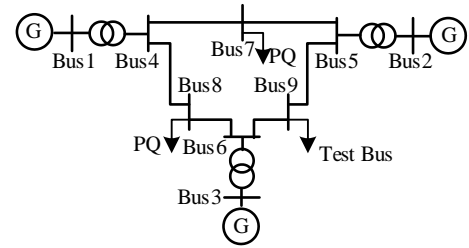

Fig. 6 Structure of WSCC 9 bus system

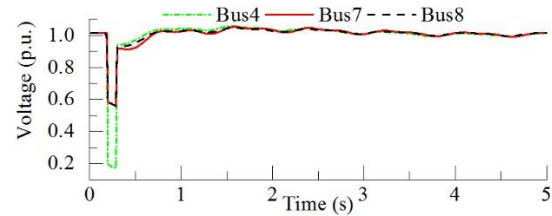

Fig. 7 Voltage curves of three faults

### B. Temporal Clustering

*1) Generation of Different Load Models*

It has been assumed in Section V.A.1 that 500 load models are identified and recorded at each load bus. In this section, the method to generate 500 different load models is introduced. In temporal clustering stage, as all the 500 different load models are identified at one load bus, it is reasonable to assume some of them are similar, which has been explained in Section III.B.

For one load bus, 10 basic load models are firstly designated. Afterwards, 500 different load models are generated around the basic load models. When generating a new load model, a basic load model is firstly chosen from 10 basic load models. Then, Gaussian noises are added to each of the parameters to generate the random changes. The standard deviation is 3% of the parameter's value. The method of load model generation is similar to that in [13], in which more details can be found.

*2) An Example of Temporal Clustering at One Load Bus*

After obtaining 500 different load models at each load bus, the next step is temporal clustering at all the load buses to pick up RLMs of each load bus. In this section, an example of temporal clustering at Bus 13 is given.

Firstly, three PFR curves under three faults for each of the 500 load models are generated in the test system, after which the distance between each two of the 500 load models can be calculated. Then, $\rho$ and $\delta$ of each load model can be calculated according to its distance to other load models, after which the decision graph can be formed, as given in Fig. 8 (a). Judging from the decision graph, the number of clusters is chosen as 10, and 10 cluster centers are picked up, which are marked in color in Fig. 8 (a). In addition, there is one outlier, which is pointed out in Fig. 8 (a). Finally, the other load models which are not cluster centers are assigned to one of the 11 clusters. A 2d multidimensional scaling figure which aims at visualizing the distance between the elements is given in Fig. 8 (b) [15]. In this way, 500 different load models are grouped into 11 clusters, and 11 cluster centers are chosen as the RLMs of Bus 13.

To validate the temporal clustering results, the PFR curves under the three-phase to ground fault at Bus7 are given in Fig. 9. Fig. 9 (a) shows the PFR curves of 11 RLMs, while Fig. 9 (b) shows the PFR curves of the 57 load models which are grouped into Cluster 1. The center of Cluster 1 and the elements in Cluster 1 are pointed out in Fig. 8 (a) and (b), respectively. From the PFR curves, it can be observed that the PFR curves of 11 RLMs are different, while those of the load models in Cluster 1 are similar. Therefore, the effectiveness of the temporal clustering is validated.

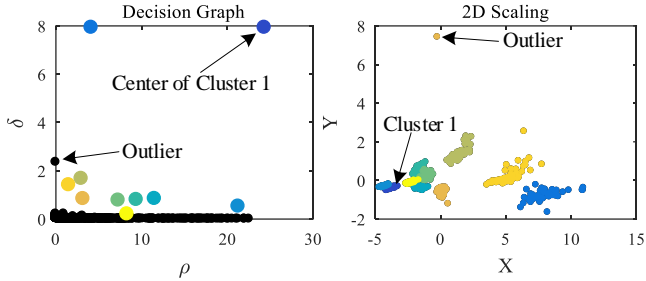

Fig. 8 Decision graph and 2D scaling figure in the temporal clustering

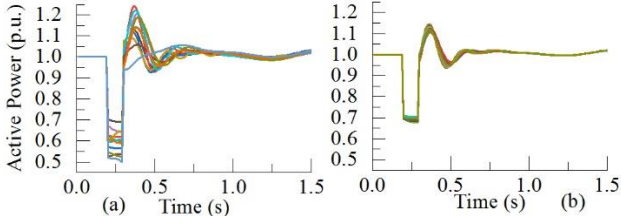

Fig. 9 Active power curves of 11 RLMs and the load models in Cluster 1

*3) Results Summary*

Similarly, temporal clustering is also conducted at other 9 load buses. The numbers of clusters in all the load buses are chosen as 10, which have matched the numbers of basic load models in load model generation. Apart from the cluster centers, 2 outliers are found in temporal clustering. Therefore, altogether 102 RLMs are sent to the control center from different load buses for further spatial clustering.

### C. Spatial Clustering

*1) Dynamic Load Models*

Firstly, the spatial clustering of dynamic load models is conducted. **RBD** is obtained from temporal clustering results first, which includes 102 elements. Then, three PFR curves under three faults for all the 102 elements in **RBD** are generated in the test system, and then the distance between each two of the elements can be calculated. Like the procedures in the previous section, the decision graph is formed and then the cluster centers are picked up. Afterwards, the other elements are assigned to one of the clusters, respectively. Three times of clustering with $nc$=3, 5, and 7 is conducted. As an example, the decision graph and the 2d scaling figure for $nc$=5 scenario are given in Fig. 10. There is one outlier in this case, which means there are 6 RLMs. In addition, there is another element that is assigned to the cluster of the outlier. The PFR curves of 6 RLMs and all the elements of Cluster 1 (as pointed out in Fig. 10) are given in Fig. 11 (a) and (b), respectively.

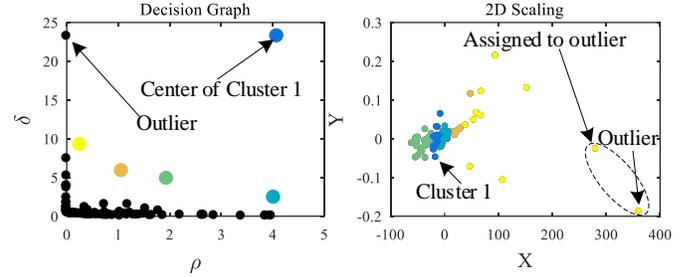

Fig. 10 Decision graph and 2D scaling figure in the spatial clustering of dynamic load models ($nc$=5)

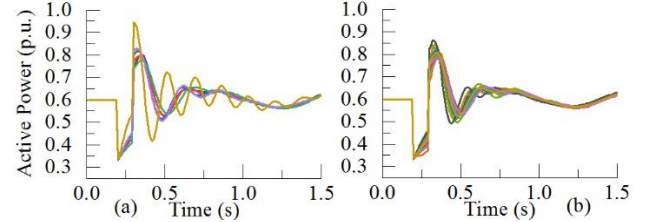

Fig. 11 Active power curves of 6 RLMs and the load models in Cluster 1

*2) Static Load Models*

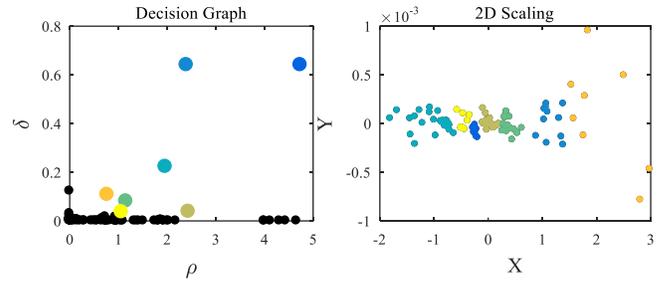

Fig. 12 Decision graph and 2D scaling figure in the spatial clustering of active static load models ($nc$=7)

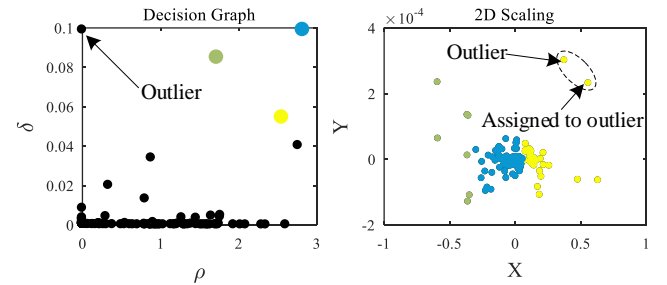

Fig. 13 Decision graph and 2D scaling figure in the spatial clustering of reactive static load models ($nc$=3)

In this section, the spatial clustering of active and reactive load models is conducted respectively. The procedures are similar to those in Section V.C.1. RBA and RBR are obtained from temporal clustering results and then RA and RR are picked up from them respectively. Also, three times of clustering is conducted with $nc$=3, 5, and 7. One example of active static load model clustering is given in Fig. 12, in which $nc$=7. There is no outlier in this case. Another example of reactive static load model clustering is given in Fig. 13, in which *nc*=3. There is one outlier and one element assigned to outlier in this case.

*3) Discussion*

From the spatial clustering results, it can be observed that the clusters are not as dense as temporal clustering results. This can also be concluded from the PFR curves in Fig. 11. This is because the load models are independently randomly generated, which is different from temporal clustering. In temporal clustering, the load models are generated around some basic load models. However, spatial clustering can still group load models with similar PFR curves into one cluster, and the load models in one cluster will have similar dynamic performance after subjected to large disturbance events. This will be validated in the following sections.

*D. Validation*

*1) Validation Scenarios*

The three-phase to ground fault at Bus 32 is simulated to validate the dynamic performance of the load models. 1000 different validation cases are simulated. In each validation case, the load models of the 10 test load buses with varying load models are randomly selected from the 500 different load models of each bus. Five different scenarios are simulated with different load models as follows:

**Scenario 1** (Ori): The original load models ($Pa_{ij}$) are used. **Scenario 2** (Tem): The load models are replaced by the RLMs of the load buses after temporal clustering ($rb_{ik}$). **Scenario 3** (Spa3): The active static load models, reactive static load models and the dynamic load models in $rb_{ik}$ are replaced by the RLMs of the system after spatial clustering ($\tilde{rb}_{ik}$). The number of clusters (*nc*) is 3. **Scenario 4** (Spa5): *nc* is changed to 5, while the other settings are the same as Scenario 3. **Scenario 5** (Spa7): *nc* is changed to 7, while the other settings are the same as Scenario 3. In this way, 5 different scenarios are simulated for one validation case. The fitting degree is defined as follows to measure the similarity of two scenarios:

$$Fitting\_Degree = 1 - \frac{\sum (y_1(t) - y_2(t))^2}{\sum (y_1(t) - \bar{y}_1(t))^2} \quad (1.8)$$

where $y_1(t)$ and $y_2(t)$ are two data series for comparison, and $\bar{y}_1(t)$ is the mean value of $y_1(t)$. If the fitting degree of two data series is close to 1, it indicates that these two data series are similar. The fitting degrees of P and Q are calculated separately, after which the average fitting degree is also calculated. In each validation case, the fitting degrees between Ori and the other four scenarios are calculated. It should be noted that the load models of all the 10 test load buses are replaced at the same time. Therefore, if one of the load models are not accurate enough, it will impact the fitting degrees of other load buses.

*2) An Example Validation Case*

TABLE I Load model parameters in different scenarios

|  | Ori | Tem | Spa7 |
|---|---|---|---|
| *p* | 0.57 | 0.54 | 0.54 |
| *Pas* | [0.23, 0.31, 0.46] | [0.23, 0.31, 0.46] | [0.37, 0.12, 0.50] |
| *Prs* | [0.16, 0.44, 0.40] | [0.16, 0.46, 0.38] | [0.25, 0.28, 0.46] |
| *Pd* | [0.72, 0.21, 0.23, 1.84] | [0.72, 0.21, 0.23, 1.80] | [0.73, 0.19, 0.21, 1.62] |

To better illustrate how the validation cases are simulated and analyzed, an example case is given in this section. The PFR curves of Bus 33 in a validation case are given as the example. The parameters in three scenarios are given in TABLE I. The P and Q PFR curves of 5 scenarios are given in Fig. 14 and Fig. 15, respectively. The fitting degrees between Ori and the other four scenarios are calculated as in TABLE II.

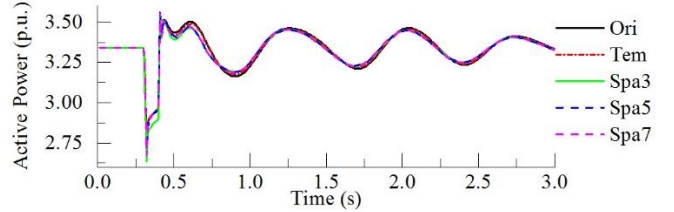

Fig. 14 Active power curves of five scenarios in the example case

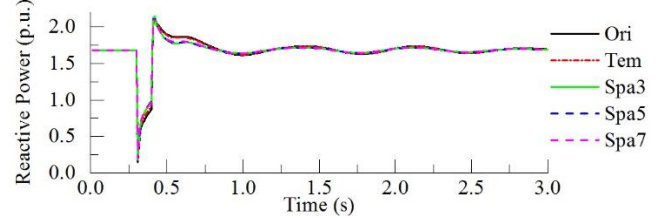

Fig. 15 Reactive power curves of five scenarios in the example case

TABLE II Fitting degrees of different scenarios in the example case

|  | Tem | Spa3 | Spa5 | Spa7 |
|---|---|---|---|---|
| FP | 0.9974 | 0.9646 | 0.9779 | 0.9769 |
| FQ | 0.9975 | 0.9763 | 0.9760 | 0.9788 |
| F | 0.9974 | 0.9704 | 0.9769 | 0.9778 |

FP: fitting degree of P, FQ: fitting degree of Q, F: average of FP and FQ

From the curves and the fitting degree results, it can be concluded that the curves of Tem are very close to the curves of Ori, while the curves of Spa3, Spa5 and Spa7 are very similar. In addition, with the increase of *nc*, the fitting degrees tend to be better, which means more accurate simulation results can be provided with more RLMs recorded. Therefore, the accuracy and effectiveness of the RLMs obtained from temporal and spatial clustering are validated in this case.

*3) Validation Results and Discussion*

TABLE III Results of fitting degrees of all the validation cases

|  | Tem | Spa3 | Spa5 | Spa7 |
|---|---|---|---|---|
| Mean(FP) | 0.9921 | 0.9068 | 0.9139 | 0.9374 |
| Mean(FQ) | 0.9929 | 0.9372 | 0.9529 | 0.9596 |
| Mean(F) | 0.9925 | 0.9220 | 0.9334 | 0.9485 |
| %(F>0.9) | 99.35 | 76.89 | 81.31 | 86.28 |
| %(F>0.95) | 98.47 | 61.11 | 68.67 | 74.49 |

FP: fitting degree of P, FQ: fitting degree of Q, F: average of FP and FQ

Similarly, the fitting degrees of 10 test load buses in all the 1000 validation cases are calculated, after which the average fitting degrees of all the validation cases are given in TABLE III. The following conclusions can be drawn from the results.

Firstly, the effectiveness of temporal clustering is validated by the high fitting degree of Tem. This can be explained from the assumption that some of the load models of one load bus identified at different time are probable to be similar. With this assumption, all the similar load models at one load bus can be represented by one RLM, through which the number of load models is reduced. From the results it can be concluded that after replacing the load models with the RLMs of the load bus, the simulation accuracy does not deteriorate much because the fitting degrees are still very close to 1.

Secondly, the effectiveness of spatial clustering is validated. The fitting degrees of the spatial clustering results are still at a high level. In addition, with the increase of *nc*, the fitting
7

degrees are also increasing. Therefore, it can be concluded that the accuracy of simulation results is still acceptable after the load models are replaced by the RLMs obtained through spatial clustering. This can be explained from two aspects. On one side, the values of *p* remain unchanged before and after spatial clustering, which can keep the proportion of dynamic load. On the other side, the models of active static load, reactive static load and dynamic load are clustered separately. All the models of thee three parts are replaced with a similar RLM, which can keep the dynamic properties as much as possible.

*E. Comparison*

*1) Storage Space*

TABLE IV Comparison of storage space for the load models of 10 load buses

|     | Number of IMPs | Number of SPs | Number of DPs | Number of indexes | Total space (bytes) |
| --- | --- | --- | --- | --- | --- |
| Ori | 4*500*10 | 6*500*10 | 500*10 | N/A | 220000 |
| Tem | 4*10*10 | 6*10*10 | 10*10 | N/A | 4400 |
| Spa | 4*nc* | 6*nc* | 10*10 | 3*10*10 | 40*nc*+700 |

The primary target of load model clustering is to process the large number of load models obtained from ambient signals-based identification and then to reduce the storage space required by load models. In this section, the storage space of the original load models (Ori), the RLMs after temporal clustering (Tem), and the RLMs after spatial clustering (Spa) is compared, the results of which are given in TABLE IV. It can be observed that through hierarchical clustering, the storage space of load model parameters is significantly reduced. Assume that each parameter is stored as a float variable (4 bytes) and each index is stored as a char variable (1 bytes). Compared with Ori, the storage space of Tem is reduced by 98%. Compared with Tem, the storage space of Spa is further reduced by 77.73% (*nc*=7).

*2) Parameters based Clustering*

TABLE V Comparison of fitting degree with parameters-based clustering

|  | Tem | Spa3 | Spa5 | Spa7 |
| --- | --- | --- | --- | --- |
| Parameter | 0.9772 | 0.9123 | 0.9251 | 0.9380 |
| PFR | 0.9925 | 0.9220 | 0.9334 | 0.9485 |

In this paper, the distance between load models is defined as the distance between PFR curves (PFR based clustering). Another possible choice is to define the distance between load models as the Euclidian distance between the vectors of load model parameters' values, which is called parameter-based clustering. In this section, the hierarchical clustering and validation procedures are re-conducted with the parameter based distance between load models, after which the average fitting degrees are compared with PFR based clustering results in TABLE V. It can be concluded that with PFR based clustering, more accurate validation results can be obtained compared with parameter-based clustering. This has validated the effectiveness of the proposed PFR based distance.

*3) K-medoids Algorithm*

K-medoids (KM) algorithm has been applied in load model clustering in previous research [13]. Compared with KM, the FDC algorithm applied in this paper has the following advantages. Firstly, the cluster centers in KM are randomly initialized and then updated through an iteration process, which means inappropriate initial cluster centers may lead to inaccurate clustering results. Secondly, the iteration process is much more time consuming compared with FDC. Thirdly, the choice of number of clusters in FDC algorithm can be made according to the decision graph, while in KM number of clusters can only be designated before clustering without any basis. Finally, KM does not have the ability to pick up outliers.

## VI. CONCLUSION

In this paper, a hierarchical clustering framework of load models is proposed. Through temporal clustering and spatial clustering, the RLMs of load buses and then the RLMs of the system are picked up. The PFR based distance of load models is defined, and the FDC algorithm is applied in load model clustering. The case study results in IEEE 39 bus system have shown that the proposed approach can successfully pick up RLMs through temporal and spatial clustering. The storage space of load models is significantly reduced without deteriorating the simulation accuracy. In addition, the proposed approach has shown its advantage over other approaches.

REFERENCE

[1] T. Han, Y. Chen, and J. Ma, "Multi-objective robust dynamic var planning in power transmission girds for improving short-term voltage stability under uncertainties," *IET Gener. Transm. Dis.,* vol. 12, no. 8, pp. 1929-1940, 2018.
[2] Y. Xu, J. Ma, Z. Y. Dong, and D. J. Hill, "Robust transient stability-constrained optimal power flow with uncertain dynamic loads," *IEEE Trans. Smart Grid,* pp. 1-11, 2016.
[3] Y. Xu, Z. Y. Dong, R. Zhang, X. Xie, and D. J. Hill, "Risk-averse multi-objective generation dispatch considering transient stability under load model uncertainty," *IET Gener. Transm. Dis.,* vol. 10, no. 11, pp. 2785-2791, Aug 4 2016.
[4] V. V, S. C. S. Srivastava, and S. Chakrabarti, "A robust decentralized wide area damping controller for wind generators and facts controllers considering load model uncertainties," *IEEE Trans. Smart Grid,* pp. 1-1, 2016.
[5] X. Wang, Y. Wang, D. Shi, J. Wang, and Z. Wang, "Two-stage wecc composite load modeling: A double deep q-learning networks approach," *IEEE Trans. Smart Grid,* pp. 1-13, 2020.
[6] M. Cui, J. Wang, Y. Wang, R. Diao, and D. Shi, "Robust time-varying synthesis load modeling in distribution networks considering voltage disturbances," *IEEE Trans. Power Syst.,* vol. 34, no. 6, pp. 4438-4450, Nov. 2019.
[7] X. Zhang, C. Lu, and Y. Wang, "A two-stage framework for ambient signal based load model parameter identification," *Int. J. Elec. Power Energy Syst.,* pp. 1-10, 2020.
[8] C. Wang, Z. Wang, J. Wang, and D. Zhao, "Robust time-varying parameter identification for composite load modeling," *IEEE Trans. Smart Grid,* vol. 10, no. 1, pp. 967-979, Jan. 2019.
[9] J. Zhao, Z. Wang, and J. Wang, "Robust time-varying load modeling for conservation voltage reduction assessment," *IEEE Trans. Smart Grid,* vol. 9, no. 4, pp. 3304-3312, July 2018.
[10] J. Ma, R. He, and D. J. Hill, "Load modeling by finding support vectors of load data from field measurements," *IEEE Trans. Power Syst.,* vol. 21, no. 2, pp. 726-735, May 2006.
[11] P. Wang, Z. Zhang, Q. Huang, N. Wang, X. Zhang, and W.-J. Lee, "Improved wind farm aggregated modeling method for large-scale power system stability studies," *IEEE Trans. Power Syst.,* vol. 33, no. 6, pp. 6332-6342, 2018.
[12] Y. Wang, Q. Chen, C. Kang, and Q. Xia, "Clustering of electricity consumption behavior dynamics toward big data applications," *IEEE Trans. Smart Grid,* vol. 7, no. 5, pp. 2437-2447, 2016.
[13] X. Zhang and D. J. Hill, "Clustering of uncertain load model parameters with k-medoids algorithm," in *Proc. 2018 IEEE PES General Meeting,* Portland, OR, 2018, pp. 1-5.
[14] A. Rodriguez and A. Laio, "Clustering by fast search and find of density peaks," *Science,* vol. 344, no. 6191, pp. 1492-1496, 2014.
[15] I. Borg and P. Groenen, *Modern multidimensional scaling: Theory and applications.* New York: Springer-Verlag, 2005.